# Using LLMs to Capture Users' Temporal Context for Recommendation

Milad Sabouri
DePaul University
Chicago, USA
msabouri@depaul.edu

Masoud Mansoury
Delft University of Technology
Delft, Netherlands
m.mansoury@tudelft.nl

Kun Lin
DePaul University
Chicago, USA
klin13@depaul.edu

Bamshad Mobasher
DePaul University
Chicago, USA
mobasher@cs.depaul.edu

## Abstract

Effective recommender systems demand dynamic user understanding, especially in complex, evolving environments. Traditional user profiling often fails to capture the nuanced, temporal contextual factors of user preferences, such as transient short-term interests and enduring long-term tastes. This paper presents an assessment of Large Language Models (LLMs) for generating semantically rich, time-aware user profiles. We do not propose a novel end-to-end recommendation architecture; instead, the core contribution is a systematic investigation into the degree of LLM effectiveness in capturing the dynamics of user context by disentangling short-term and long-term preferences. This approach, framing temporal preferences as dynamic user contexts for recommendations, adaptively fuses these distinct contextual components into comprehensive user embeddings. The evaluation across Movies&TV and Video Games domains suggests that while LLM-generated profiles offer semantic depth and temporal structure, their effectiveness for context-aware recommendations is notably contingent on the richness of user interaction histories. Significant gains are observed in dense domains (e.g., Movies&TV), whereas improvements are less pronounced in sparse environments (e.g., Video Games). This work highlights LLMs' nuanced potential in enhancing user profiling for adaptive, context-aware recommendations, emphasizing the critical role of dataset characteristics for practical applicability.

## 1 Introduction

Accurate modeling of user preferences is essential for effective recommender systems, particularly in dynamic environments where user interests continually evolve. Context-Aware Recommender Systems (CARS) traditionally leverage predefined, static contextual factors, which limits their ability to capture nuanced temporal dynamics such as transient short-term interests versus persistent long-term tastes. This represents a critical gap in existing methods for dynamically modeling user contexts. Recent advancements in Large Language Models (LLMs) offer promising capabilities for semantically rich, context-aware user profiling. However, current LLM-based profiling methods often produce singular, static summaries of user histories, neglecting the differentiation between short-term and long-term preferences. This research specifically addresses this gap by systematically evaluating the effectiveness of LLMs in capturing temporal user dynamics. Rather than introducing a novel end-to-end recommendation architecture, the primary contribution is a controlled investigation of how effectively LLMs can distinguish and fuse natural language summaries generated separately from recent (short-term) and historical (long-term) user interactions, resulting in comprehensive, temporally-aware user embeddings. We assess this approach across two distinct domains, Movies&TV and Video Games, highlighting that the effectiveness of LLM-driven temporal profiling significantly depends on the density and dynamism of user interaction histories. Our observations indicate pronounced performance improvements in domains characterized by dense interactions (e.g., Movies&TV), while the benefits are less significant in sparser environments (e.g., Video Games). These insights critically inform the practical applicability and conditional effectiveness of leveraging LLMs for enhanced dynamic user representation in context-aware recommendation settings.

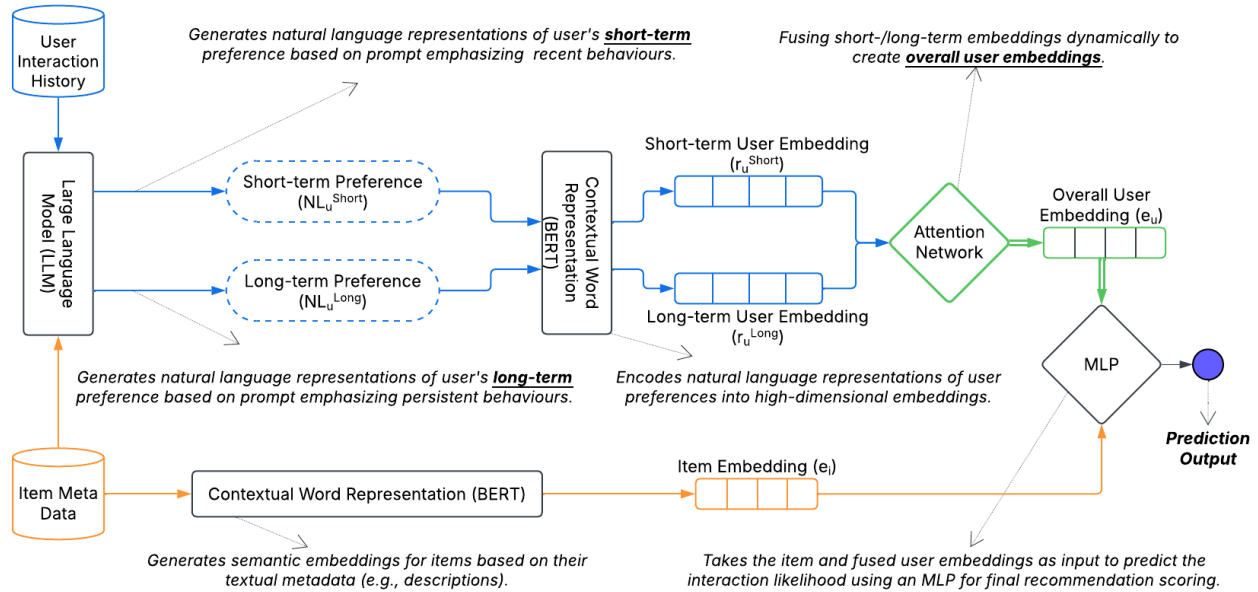

Figure 1: Architectural Framework for Assessing LLM-Driven Dynamic User Context Profiling

## 2 Related Works

Early recommender systems predominantly employed collaborative filtering [9, 15] or content-based methods, which typically distilled item metadata into fixed-length vector representations. The emergence of deep language models, like BERT [4] and advanced Large Language Models (LLMs) [21, 28], significantly enhanced content-based techniques [25]. These advancements enabled more nuanced semantic understanding of user interaction histories, inferring richer user profiles [10, 27] beyond simplistic features. The field of Context-Aware Recommender Systems (CARS) [2, 11, 12] explicitly addresses the need to tailor recommendations based on varying contextual factors. Within CARS, a key direction has focused on explicitly differentiating between long-term and short-term user preferences, recognizing time as a critical context dimension. Recent LLM-based user modeling research also largely focuses on user profile generation. Some approaches derive compact user embeddings from temporally rich histories [14] or enrich user-item graphs with LLM-generated attributes [24]. Others demonstrate LLMs encoding temporal recency via prompt engineering [3] or fine-tuning for instruction-following recommendations [7]. While confirming LLMs' efficacy in capturing rich semantic contexts, these methods often treat user profiles as singular, static summaries, neglecting dynamic contextual nuances and the crucial distinction between transient short-term interests and stable long-term preferences. A systematic assessment of their efficacy across varying temporal granularities and data characteristics is limited. Complementary research on dual-timescale models in CARS has shown the value of dynamically integrating temporal scales through sophisticated architectures [26], including spatial recommendation systems for next-point-of-interest predictions [22]. However, these typically rely on conventional embedding methods or sequence-based encoders [5, 8, 20], and do not fully exploit the advanced semantic richness of LLM-generated representations. Furthermore, their fusion mechanisms [19] often lack adaptive attention explicitly conditioned on the user's current context, limiting their dynamic adaptability for true context-aware solutions. Despite the independent promise of LLM-based user profiling and dual-timescale modeling for user dynamics, their comprehensive and systematic integration remains largely under-explored for context-aware applications. Existing work seldom examines how LLMs can enhance the semantic abstraction of user preferences within explicitly segmented temporal frameworks, nor provides a comprehensive evaluation of this integrated approach across diverse user behavior patterns and interaction densities.

This work directly addresses this critical intersection, advancing the field of CARS. We introduce a modular architecture for controlled evaluation to systematically assess the degree of effectiveness of LLM-driven temporal user profiling.. Unlike prior LLM-based profile-generation approaches [7, 14, 24] that often produce singular summaries, the provided framework in this paper innovatively generates separate, semantically rich short-term and long-term profiles from raw interaction data. These dual contextual narratives are then adaptively combined via an attention-based fusion layer. By focusing on this specific profiling component and evaluating it across diverse content domains, this work provides unique insights into LLMs' effectiveness for capturing nuanced, time-aware user preferences and, critically, elucidates the conditions under which these LLM-driven contextual profiles offer the most significant advantages for context-aware recommendation.

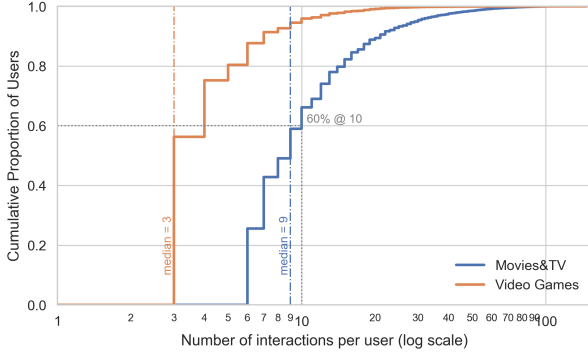

Figure 2: Empirical Cumulative Distribution Function (ECDF) of user interaction frequency for Movies&TV and Video Games datasets. This plot visually distinguishes differing user activity levels, providing critical context for assessing LLM-driven temporal user profiling under varying historical data richness.

## 3 Methodology

This study assesses LLM-driven user profiling within dynamic user preferences. Users $u \in \mathcal{U}$ interact with items $i \in \mathcal{I}$, recording chronological history $\mathcal{H}_u = \{(i, t_{u,i}) : i \in \mathcal{I}_u\}$. The objective is to learn user $\mathbf{e}_u$ and item $\mathbf{e}_i$ embeddings to predict interaction probability via a function $f : \mathbb{R}^d \times \mathbb{R}^d \to [0, 1]$:

$$\hat{y}_{u,i} = f(\mathbf{e}_u, \mathbf{e}_i), \tag{1}$$

where $\hat{y}_{u,i}$ is the predicted likelihood, implemented via a multi-layer perceptron (MLP).

The architectural framework (Figure 1) serves as a modular evaluation tool for assessing LLM-driven user profiling, integrating *temporal dynamics* and *semantic information* as core contextual elements. Its purpose is to analyze these contextual components in user profile generation. It comprises: (i) Generation of Temporal User Profiles (contextual representation), (ii) Semantic Embedding Representation, and (iii) Personalized Recommendation Generation.

### 3.1 Generating Temporal User Contexts via LLMs

To capture evolving *user context*, we leverage Large Language Models (LLMs) to generate two distinct natural language user profiles: short-term and long-term. Given a user's chronological history $\mathcal{H}_u$, the LLM creates separate summaries:

*Short-Term Contextual Profile:* This profile emphasizes recent interactions, capturing current and transient user interests as a short-term *context*. For instance, the prompt guides the LLM to "Summarize the user's most recent interactions (e.g., last $N$ items) focusing on immediate interests and temporary trends." This is formalized as:

$$\mathrm{NL}_u^{\mathrm{short}} = \mathrm{LLM}(\mathcal{H}_u, \mathrm{Prompt}^{\mathrm{short}}). \tag{2}$$

*Long-Term Contextual Profile:* This profile considers full historical interactions, capturing stable, enduring long-term preferences as a persistent *context*. An example prompt is "Analyze the user's entire interaction history to identify stable, persistent preferences and overarching themes." This is formalized as:

$$\mathrm{NL}_u^{\mathrm{long}} = \mathrm{LLM}(\mathcal{H}_u, \mathrm{Prompt}^{\mathrm{long}}). \tag{3}$$

This dual prompting enables distinct *contextual representations* differentiating recent from persistent preferences. All Prompts are available in our GitHub repository (Section 4).

### 3.2 Semantic Embedding Representation

Natural language user profiles ($\mathrm{NL}_u^{\mathrm{short}}$ and $\mathrm{NL}_u^{\mathrm{long}}$) are transformed into dense, semantically meaningful vector spaces using a pre-trained BERT [4] model:

$$\mathbf{r}_u^{\mathrm{short}} = \mathrm{BERT}(\mathrm{NL}_u^{\mathrm{short}}), \tag{4}$$

$$\mathbf{r}_u^{\mathrm{long}} = \mathrm{BERT}(\mathrm{NL}_u^{\mathrm{long}}), \tag{5}$$

where $\mathbf{r}_u^{\mathrm{short}}, \mathbf{r}_u^{\mathrm{long}} \in \mathbb{R}^d$ represent distinct short-term and long-term semantic profile embeddings. Item embeddings are similarly derived from textual descriptions using the same BERT encoder, ensuring consistency.

To form a comprehensive final user profile embedding ($\mathbf{e}_u$), an attention mechanism [23] dynamically fuses short-term and long-term embeddings. This adaptive fusion is crucial as relative importance of different *temporal contexts* varies across users. The attention weights are computed as follows:

$$\alpha_u^{\mathrm{short}} = \frac{\exp(\mathbf{W}_a \mathbf{r}_u^{\mathrm{short}})}{\exp(\mathbf{W}_a \mathbf{r}_u^{\mathrm{short}}) + \exp(\mathbf{W}_a \mathbf{r}_u^{\mathrm{long}})}, \tag{6}$$

$$\alpha_u^{\mathrm{long}} = 1 - \alpha_u^{\mathrm{short}}, \tag{7}$$

where $\mathbf{W}_a \in \mathbb{R}^{1 \times d}$ is a learnable weight matrix. The final fused user embedding $\mathbf{e}_u$ is then computed as:

$$\mathbf{e}_u = \alpha_u^{\mathrm{short}} \cdot \mathbf{r}_u^{\mathrm{short}} + \alpha_u^{\mathrm{long}} \cdot \mathbf{r}_u^{\mathrm{long}}, \tag{8}$$

effectively allowing the model to adaptively balance temporal preferences. The efficacy of this attention-based fusion mechanism is further validated in our ablation study (Section 5).

### 3.3 Personalized Recommendation Generation

To predict user-item interaction likelihood, the combined user embedding $\mathbf{e}_u$ (from LLM-driven temporal profiling) and item embedding $\mathbf{e}_i$ are concatenated and processed by an MLP:

$$\hat{y}_{u,i} = \mathrm{MLP}([\mathbf{e}_u; \mathbf{e}_i]), \tag{9}$$

where $[\mathbf{e}_u; \mathbf{e}_i] \in \mathbb{R}^{2d}$ is the concatenated input. The MLP uses ReLU activations [13] and a sigmoid output for $[0, 1]$ probabilities.

Model parameters are optimized using binary cross-entropy loss:

$$\mathcal{L} = -\frac{1}{|\mathcal{D}|} \sum_{(u,i,y_{u,i}) \in \mathcal{D}} \left( y_{u,i} \log \hat{y}_{u,i} + (1 - y_{u,i}) \log(1 - \hat{y}_{u,i}) \right), \tag{10}$$

where $\mathcal{D}$ is the training set, and $y_{u,i}$ indicates actual interaction.

Table 1: Datasets and User Profiles Statistics

| Dataset   | Users  | Items  | Interactions | Mean  | Median | Mode | Std Dev |
|-----------|--------|--------|--------------|-------|--------|------|---------|
| Movies&TV | 10,000 | 14,420 | 202,583      | 11.79 | 9.00   | 6    | 9.80    |
| Games     | 10,371 | 3,790  | 83,842       | 4.55  | 3.00   | 3    | 3.97    |

The LLM-driven temporal profiles enhance recommendation performance for **context-aware systems** through two key elements. First, **temporal awareness** explicitly distinguishes and dynamically fuses short-term and long-term user preferences, capturing user behavior as a **dynamic context**. Second, **semantic richness** leverages natural language representations for precise user-item alignment within diverse item **contexts**. These design choices improve accuracy in dynamic, semantically rich environments, representing complex contextual landscapes. The experimental assessment (subsequent sections) examines each profiling component's contribution and conditional effectiveness.

## 4 Experiments

For benchmarking, we compare against four baselines: **Centric** [16], a content-based approach without modeling *temporal context*; **Temp-Fusion**, which is designed to assess the benefits of temporal fusion without LLM-generated semantic profiles by segmenting and fusing short-term/long-term numerical item embeddings, thus isolating the LLM's contribution to *semantic contextualization*; and two control baselines, **Popularity** [18] (a non-personalized ranking method lacking contextual modeling) and **Matrix Factorization (MF)** [9] (a collaborative filtering technique without textual features or explicit contextual signals). Evaluation employs a rigorous temporal holdout (60% train, 20% validation, 20% test), crucial for assessing prediction in evolving user *contexts*. Performance is reported using Recall@K and NDCG@K.

For the LLM-driven approach, user profiles were generated using GPT-4o-mini [1], with textual summaries encoded via SBERT [17] (MiniLM-L6-v2, 384-dim). Interaction likelihoods are predicted using an MLP (hidden size 128, dropout 0.2) trained with binary cross-entropy, a batch size of 2048, and the Adam optimizer. Training incorporates early stopping (patience=5) for up to 100 epochs, executed on four NVIDIA A100 GPUs. All code, data, and prompts are publicly available[1].

### 4.1 Datasets and Domain Characteristics

The LLM-driven temporal user profiling framework is evaluated on two publicly available datasets from the Amazon Product Reviews corpus [6]: *Movies&TV* and *Video Games*. These datasets were filtered for English item descriptions exceeding 500 characters to ensure semantically rich input. Domains were chosen to assess the profiling strategy under **divergent user behavior patterns and data richness**, representing distinct *contextual landscapes* influencing temporal dynamics. Table 1, and Figure 2 summarize the contrasting dataset characteristics crucial for our study. The *Movies&TV* dataset is considerably larger (14,420 unique items, 202,583 interactions), with higher average user activity (mean 11.79 interactions per user, median 9.00) and high sparsity (approx. 0.9986). This offers rich, diverse user histories, ideal for evaluating LLM semantic depth in dense *contextual signals*. In contrast, the *Video Games* dataset is smaller (3,790 unique items, 83,842 interactions) with lower per-user activity (mean 4.55, median 3.00) and slightly lower sparsity (approx. 0.9979), presenting a challenging *sparse contextual environment* where recent trends may play a more dominant role. This selection enables assessment of the LLM-driven method's ability to capture nuanced preferences from diverse histories and its robustness in sparser, less dynamic *contextual environments*. Chronological sorting of rating reviews creates a valid temporal signal, enabling differentiation of short-term versus long-term user preferences. This approach helps understand LLMs' conditional *contextual effectiveness* and practical applicability in temporal user profiling across varying data densities.

### 4.2 Results and Discussion

The experimental evaluation assessed the LLM-based temporal user profiling (LLM-TP) method for context-aware recommendation across two distinct datasets. Table 2 presents comprehensive performance via Recall@K and NDCG@K. Baseline performance revealed non-personalized Popularity as a lower bound. On Movies&TV, Popularity slightly surpassed Matrix Factorization (MF), indicating MF's challenge in sparse datasets to discern patterns from interaction matrices in such *contexts*. Centroid-Based (Centric), leveraging content embeddings, outperformed Popularity and MF, highlighting the value of *content-based context*. Temporal Fusion (Temp-Fusion), fusing short-term and long-term numerical item embeddings via attention, further improved over Centric, demonstrating the significant value of capturing *dynamic user context*, particularly for Video Games. The LLM-TP method achieved the highest overall performance, consistently surpassing all baselines. This statistically significant superiority over Centric underscores the efficacy of combining temporal user modeling with rich, semantically-aware LLM representations to generate powerful *contextual representations*. A crucial nuance emerged regarding LLM-driven profiling efficacy. While substantial gains occurred on Movies&TV, Video Games performance was more complex: Temp-Fusion slightly led on Recall@10, NDCG@10, and NDCG@20, while LLM-TP maintained an edge in Recall@20. This suggests that in sparser domains, direct aggregation might better optimize top-K ranking, whereas LLMs excel at capturing broader semantic themes for wider recall. The larger gain on Movies&TV indicates LLMs' richer semantic understanding is most impactful in heterogeneous, activity-rich *contextual environments*.

Conditional effectiveness raises deployment considerations for context-aware systems. LLMs' computational cost is best justified in contexts with rich, dynamic user histories (e.g., Movies&TV), where

---
[1] https://github.com/milsab/LLM-TP

Table 2: Comparative Performance Evaluation of LLM-Driven Temporal User Profiling Across Different Domains

| Method | Movies&TV | | | | Video Games | | | |
|---|---|---|---|---|---|---|---|---|
| | Recall@10 | NDCG@10 | Recall@20 | NDCG@20 | Recall@10 | NDCG@10 | Recall@20 | NDCG@20 |
| Centric | 0.0113 | 0.0191 | 0.0199 | 0.0269 | 0.0645 | 0.0532 | 0.0932 | 0.0649 |
| Popularity | 0.0082 | 0.0145 | 0.0133 | 0.0191 | 0.0397 | 0.0324 | 0.0706 | 0.0453 |
| MF | 0.0048 | 0.0085 | 0.0087 | 0.0124 | 0.0457 | 0.0370 | 0.0754 | 0.0491 |
| Temp-Fusion | 0.0118 | 0.0201 | 0.0207 | 0.0276 | **0.0693** | **0.0589** | 0.0982 | **0.0712** |
| **LLM-TP** | **0.0132*** | **0.0217*** | **0.0223*** | **0.0293*** | 0.0665* | 0.0547* | **0.1021*** | 0.0683* |
| Gain of LLM-TP vs. Centric | 17% | 14% | 12% | 9% | 3% | 3% | 10% | 5% |

*asterisk indicates that the improvement of the proposed method over the baseline (Centric) is statistically significant ($p < 0.05$).*

nuanced capture yields substantial performance. In sparse environments, benefits may be marginal compared to simpler aggregation, warranting cost-benefit analysis.

In conclusion, the results demonstrate the compelling advantage of incorporating temporal dynamics and LLM-based semantic representations into user profiling for context-aware recommendation. LLM-TP's consistent improvements highlight LLMs' added value. Crucially, differential gains across datasets underscore considering dataset characteristics (sparsity, catalog size, user activity patterns) when designing and deploying LLM-driven profiling approaches in diverse *contextual landscapes*.

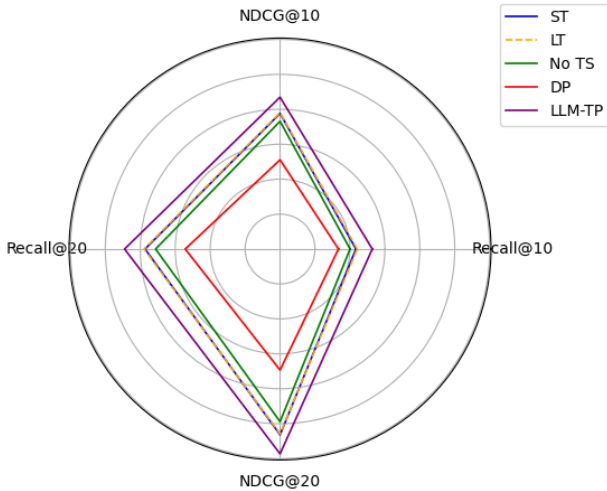

Figure 3: Performance comparison of the full LLM-TP method against its ablated variants on the Movies&TV dataset. This radar plot visually demonstrates each component's critical and synergistic contribution to LLM-based user profiling for context-aware recommendations across key metrics.

## 5 Ablation Study

This ablation study analyzes the individual contributions of LLM-based semantic summarization, temporal context separation, and adaptive fusion to the user profiling framework's overall effectiveness. Conducted primarily on the *Movies&TV* domain due to its richer *contextual signals*, results for *Video Games* consistently followed the same performance ordering (omitted for brevity).

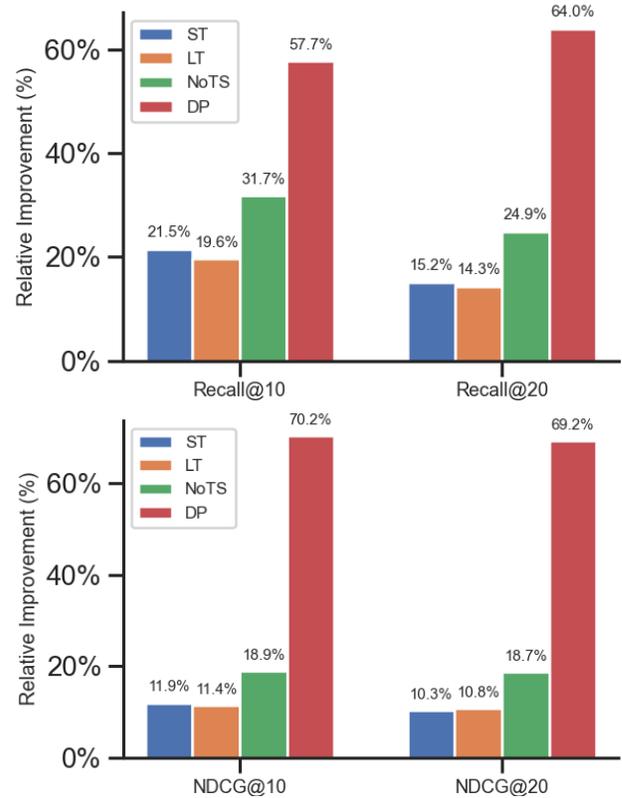

Figure 4: Relative performance gains (%) of the full LLM-TP method over its ablated variants on the Movies&TV dataset.

The study involved four distinct variants designed to isolate specific architectural impacts: *Short-Term Preferences Only (ST)* and *Long-Term Preferences Only (LT)* assessed the isolated impact of immediate and enduring user interests (temporal contexts), respectively. *General Preferences without Temporal Distinction (NoTS)* removed the temporal split, collapsing all interactions into a single textual summary to examine the necessity of explicit temporal context segmentation. Finally, *Substitute MLP with Dot Product (DP)* maintained LLM-generated temporal user embeddings but replaced the non-linear MLP scoring function with a simpler dot product, quantifying the prediction layer's role in leveraging rich LLM-generated *contextual embeddings*. Analyzing performance (Figure 3, Tables 3) reveals each component's critical role. The DP variant, innermost in Figure 3, resulted in the most substantial performance degradation, underscoring the MLP's importance for effectively utilizing rich LLM-generated embeddings. Temporal modeling variants unequivocally highlight the importance of distinguishing short-term and long-term preferences for robust user profiling. NoTS, discarding temporal separation, consistently performed worse than the full method (Fig. 3; relative gains exceeding 20% for Recall and 19% for NDCG in Fig. 4), empirically validating explicit temporal context modeling. While ST and LT variants positively contribute individually (polygons enclosed by full model in Fig. 3), their synergistic combination and adaptive fusion within the full model are necessary for optimal performance ( 21.5% Recall@10 over ST, 19.6% over LT in Fig. 4).

Table 3: Performance of LLM-Driven Profiling Ablations on Movies&TV

| Experiment | Recall@10 | Recall@20 | NDCG@10 | NDCG@20 |
|---|---|---|---|---|
| Short-Term Only (ST) | 0.0109 | 0.0193 | 0.0194 | 0.0266 |
| Long-Term Only (LT) | 0.0110 | 0.0195 | 0.0195 | 0.0265 |
| General Preferences (No TS) | 0.0100 | 0.0178 | 0.0183 | 0.0247 |
| Dot-Product Scoring (DP) | 0.0084 | 0.0136 | 0.0128 | 0.0173 |
| **LLM-TP** | **0.0132** | **0.0223** | **0.0217** | **0.0293** |

The ablation study conclusively demonstrates that the LLM-based temporal user profiling method's superior performance stems from the synergistic combination of its key components. Explicitly modeling both short-term and long-term user preferences is vital, confirmed by the significant NoTS performance drop. Both immediate (ST) and enduring (LT) preferences contribute positively. Moreover, the non-linear MLP scoring mechanism plays a pivotal role, enabling the model to effectively leverage the learned user and item *contextual representations*. Thus, the study validates the LLM-driven profiling design, confirming its strength in integrated temporal profiling and effective embedding interaction through a non-linear scoring function.

## 6 Conclusion

This research provides a comprehensive assessment of LLM-based temporal user profiling for context-aware recommendation. The provided modular framework explicitly disentangles short-term and long-term user preferences, addressing limitations of conventional aggregation by providing a more nuanced user representation. Experiments demonstrate the LLM-driven profiling consistently outperforms baselines, with benefits particularly pronounced in dense, dynamic interaction environments (e.g., Movies&TV) where LLMs' deep semantic understanding excels. Conversely, gains are less significant in sparser contexts (e.g., Video Games). Ablation studies validate the critical roles of temporal separation, semantic summarization, and adaptive fusion. These observations highlight the compelling advantage of incorporating evolving and persistent user behavior via LLM-driven profiles for advancing personalized recommendation systems, underscoring the vital importance of considering dataset characteristics for practical applicability within diverse contextual landscapes. Future work will explore extending this framework to real-time contextual adaptation and addressing deployment challenges like fairness and transparency in CARS.